\documentclass[12pt]{JHEP}
\usepackage{amssymb} 

\newcommand{\Tr}{\mathop{\rm Tr}\nolimits}

\def\bra#1{\langle #1 |}
\def\ket#1{|#1 \rangle}
\def\aver#1{\langle\, #1 \,\rangle}

\def \be {\begin{equation}}
\def \ee {\end{equation}}
\def \bea {\begin{eqnarray}}
\def \eea {\end{eqnarray}}

\def \nn {{\mathbb N}}
\def \zz {{\mathbb Z}}
\def \cc {{\mathbb C}}
\def \rr {{\mathbb R}}

\def \ii {{\cal I}}

\def \aa {{\cal A}}

\def \nc {noncommutative }
\def \ncg {noncommutative geometry }
\def \sf  {string field }
\def \sft {string field theory }
\def \da {\dagger}

\title{String field theory at large B-field and noncommutative geometry}
\author{Martin Schnabl\\
{Scuola Internazionale Superiore di Studi Avanzati} \\
{Via Beirut 4, 34014 Trieste, Italy} \\
INFN, Sezione di Trieste \\ 
E-mail: {\tt schnabl@sissa.it}}

\abstract{
In the search for the exact minimum of the tachyon potential in the Witten's
cubic string field theory we try to learn as much as possible from the 
string field theory in the large B-field background. We offer a simple alternative proof of
the Witten's factorization, carry out the analysis of string field equations 
also for the noncommutative torus
and find some novel relations to the algebraic K-theory. We note an
intriguing relation between Chern-Simons and Chern classes of two
noncommutative bundles. Finally we observe a certain pattern which enables us to 
make a plausible conjecture about the exact form of the minimum.}  

\preprint{SISSA 96/2000/EP\\  
{\tt hep-th/0010034}}

\begin{document}

\section{Introduction}

One of the mysteries of string theory which has not been so far fully
understood mathematically is the Sen's conjecture
\cite{Sen:Descent}. According
to it there are solitonic solutions to string field theory
equations of motion for which the action attains rather special
values which coincide with the mass differences between various branes known
from different considerations. The conjecture was tested
with amazing precision \cite{SZ,MT} using the numerical method 
developed in \cite{Kostelecky}. Only very recently it has been proved
analytically in the framework of the so called background independent \sft
\cite{bisft1,bisft2,bisft3}. 

In spite of this recent progress the fact that the Witten-Chern-Simons \sft
\cite{Witten:NCGSFT} has this property still remains rather miraculous and lacks
understanding. One is tempted to believe that it is the \nc topological nature 
of the action which may help to prove the Sen's conjecture in the realm of this 
theory. Life is however not too easy. The main formal obstacle
to use the methods of \ncg \cite{Connes} appears to be the lack 
of any satisfactory definition of the \sf algebra which is both 
closed and associative. One may also question whether it is the right
strategy to take since an analogous conjecture is believed to be true also for
the superstring \cite{Berk1,Berk2,DeSmet,Iqbal} where no \nc geometric 
formulation is known.

Regarding the study of lower dimensional branes
there were two basic approaches in the literature. In
the first one \cite{HK,deMelloKoch, MSZ} they were studied numerically as solitons 
in the \sf theory. A systematic numerical method --- modified level expansion scheme --- 
was developed in \cite{MSZ}.
The second approach \cite{GMS, DMR, HKLM} was based on the observation
that in the large B-field limit the effective action for the
tachyon field admits simple solutions which 
can be interpreted as lower dimensional branes whose tensions exactly 
reproduce the known results.
This last series of results was beautifully
put to the string field theory setting by Witten \cite{Witten:NCtach}, who showed the
factorization of the algebra in the large B-field limit.
The beauty of his work lies mainly in the fact that two hitherto
unrelated occurrences of \ncg in string theory were shown to be
intimately connected.

We feel therefore that a good starting point to understand the mysteries
of the \sf algebra and action is to start by examining
it carefully in the limit of large B-field. In the course of this work 
we will try to pay special attention to the various \nc geometric aspects.
It should not be perhaps surprising that algebraic K-theory will show up
to play some role since it is one of the key building blocks of \ncg in its
abstract setting. The role we find here is however quite different from
the one recently discussed in a similar context in \cite{HarveyMoore,Matsuo}. 
They study the K-theory along the worldvolume of the brane whereas we focus
on the transverse directions. The K-theory we discuss is the one which 
previously appeared in \cite{CDS}. For a friendly introduction to K-theory see 
\cite{Wegge-Olsen} and for the overview of applications to string theory see
\cite{OlsenSzabo,Witten:Koverview}.

The paper is organized as follows. Section 2 is devoted to the
rederivation of the Witten's factorization in the operator language 
which is much simpler and more transparent.
The basic ingredient here is the three string
vertex in the B-field background constructed in \cite{Sugino, Kawano}. In
section 3 we observe natural correspondence between D-brane
decays and algebraic K-theory. We deal with two particular cases. In the first one 
the B-field is put on the plane $\rr^2$. Here we can utilize the GMS \cite{GMS} construction
of the projectors. In the second case we consider the B-field on a two torus ${\mathbb T}^2$ where 
it becomes necessary to use the Powers--Rieffel projectors \cite{Rieffel} or some variant thereof.
We find quite an intriguing connection between invariants of the \nc bundle over the torus 
defined by the choice of a projector and \nc bundle defined by the choice of a \sf connection.
Section 4 contains observation about mutual relations between various solutions to
\sf equations. It formally looks as a gauge transformation, but the relevant isometry is
nonunitary. This leads at the end to a plausible proposal for the exact minimum.
Concrete examples and numerical tests are postponed to the future work.

\section{String field algebra at nonzero B-field}

All the degrees of freedom of string field theory are contained in the string field 
\begin{equation}
\Psi = \int d^{26} p \, (t(p) c_1 + A_\mu(p) \alpha_{-1}^\mu c_1 + \cdots) \ket{0,p}
\end{equation}
which is an element of the Fock space of the first quantized string theory.
It is governed by the Chern-Simons type of an action
\begin{equation}\label{action}
S[\Psi]=-\frac{1}{\alpha' G_o^2} \left( \frac 12 \aver{\Psi, Q \Psi} + \frac 13 \aver{\Psi,\Psi*\Psi} \right)
\end{equation}
The noncommutative star multiplication originally defined in terms of gluing of strings 
\cite{Witten:NCGSFT} was formulated in the operator language in 
\cite{GrossJevicki,Samuel:1986}
and the B-field was taken into account in \cite{Sugino,Kawano}. 
It is defined through 
\begin{equation}
\Psi_1 * \Psi_2 = bpz \left( \bra{V} \Psi_1 \otimes \Psi_2 \right)
\end{equation}
where $bpz$ denotes the $bpz$ conjugation in conformal field theory and
\begin{eqnarray}\label{vertex} 
\bra{V} &=& \left(\frac{3\sqrt{3}}{4} \right)^3
\delta(p^{(1)}+p^{(2)}+p^{(3)}) \bra{\tilde 0} \otimes  \bra{\tilde 0} \otimes
\bra{\tilde 0}\times \\
&&
\times\exp \left( \sum_{m,n=0}^{\infty} 
 \frac 12 \alpha_n^{(r)\mu} N_{nm}^{rs} \alpha_m^{(s)\nu} G_{\mu\nu} + 
 \sum_{m=0,n=1}^{\infty} c_n^{(r)} X_{nm}^{rs} b_m^{(s)} -
 \frac{i}{2} \theta_{\mu\nu} p^{(1)\mu} p^{(2)\nu} \right)
\nonumber
\end{eqnarray}
The Neumann coefficients $N_{nm}^{rs}, X_{nm}^{rs}$ are reviewed in \cite{Taylor}. 
The vacuum $\bra{\tilde 0}$ is related to the standard $Sl(2,\rr)$ invariant vacuum $\bra{0}$ through 
$\bra{\tilde 0}= \bra{0} c_{-1}^{(i)} c_0^{(i)}$ where $i=1,2,3$ label one of the three Fock spaces 
in the tensor product. As usual $\alpha_0^\mu=\sqrt{2\alpha'} p^\mu$. 

The effective open string coupling constant, open string metric and the noncommutativity parameter 
\cite{SW} are given by 
\begin{eqnarray}
G_o &=& g_o \left(\frac{\det G}{\det(g+2\pi\alpha' B)}\right)^\frac 14 \\
G_{\mu\nu}&=&g_{\mu\nu} - (2\pi\alpha')^2 (B g^{-1} B)_{\mu\nu} \\
\theta^{\mu\nu}&=& -(2\pi\alpha')^2 \left(\frac{1}{g+2\pi\alpha' B} B \frac{1}{g-2\pi\alpha' B} 
\right)^{\mu\nu}
\end{eqnarray}
These effective parameters also appear in the formula for Virasoro generators 
(and therefore in the BRST charge $Q$) and
in the commutation relations for the Fock space generators
\begin{eqnarray}\label{commutators} 
[\alpha_m^\mu, \alpha_n^\nu] &=& m \delta_{m+n,0} G^{\mu\nu} \nonumber\\
{[x^\mu,x^\nu]} &=& \theta^{\mu\nu} \nonumber\\
{[p^\mu,x^\nu]} &=& -i\,  G^{\mu\nu} 
\end{eqnarray}

\subsection*{Large B-field limit}

Now take the limit $B \to \infty$ keeping fixed all closed string parameters (including the open string 
coupling constant $g_o$ but not the effective one $G_o$). To make things more transparent set $B=t B_0$
and take $t \to \infty$ as in \cite{Witten:NCtach}.
The effective parameters clearly depend on $t$ as 
\begin{eqnarray}
G_o &\sim& G_{o\,0} t^{r/2} \nonumber\\
G^{\mu\nu}&\sim& G_0^{\mu\nu} t^{-2}  \nonumber\\
\theta^{\mu\nu}&\sim& \theta_0^{\mu\nu} t^{-1}
\end{eqnarray}
where $r$ denotes the rank of the B-field and for the brevity let us assume that it is maximal.
Altogether the $t$ dependence enters at two places: First
in the commutation relations (\ref{commutators}) for the Hilbert space operators 
and then also explicitly in the definition (\ref{vertex}) of the star product. 

To see the change in the structure of the string field algebra we have to rescale all Fock space 
operators in such a way that their commutation relations don't depend on $t$. (We are then sure that 
we are studying different star products on the same space).
The rescaling which does that is
\begin{eqnarray}
\alpha_m^\mu &\to& \tilde{\alpha}_m^\mu = t \, \alpha_m^\mu  \qquad (m \ne 0) \nonumber\\
p^\mu &\to& \tilde{p}^\mu = t^{3/2} p^\mu  \nonumber\\
x^\mu &\to& \tilde{x}^\mu = t^{1/2} x^\mu  
\end{eqnarray}
After this rescaling the exponent in the vertex (\ref{vertex}) takes the simple form
\begin{eqnarray}
&& \sum_{m,n=1}^{\infty} \frac 12 \tilde{\alpha}_n^{(r)\mu} N_{nm}^{rs} \tilde{\alpha}_m^{(s)\nu} 
G_{0 \mu\nu} + 
   \frac{1}{\sqrt{t}} \sum_{n=1}^{\infty} \sqrt{\frac{\alpha'}{2}} \tilde{\alpha}_n^{(r)\mu} 
(N_{n0}^{rs}+N_{0n}^{rs}) \tilde{p}^{(s)\nu}  G_{0 \mu\nu}+
\nonumber\\
&&  +\frac{1}{t}\alpha' \tilde{p}^{(r)\mu} N_{00}^{rs} \tilde{p}^{(s)\nu} G_{0 \mu\nu} 
 + \sum_{m=0,n=1}^{\infty} c_n^{(r)} X_{nm}^{rs} b_m^{(s)} -
 \frac i2 \theta_{0 \mu\nu} \tilde{p}^{(1)\mu} \tilde{p}^{(2)\nu}
\end{eqnarray}
We see that in the large $t$ limit the terms which couple $\alpha$ oscillators with momenta $p$ 
vanish but the whole star product nevertheless remains nontrivial. Now the generic string field 
is the sum of terms
\begin{equation}
a \, e^{i k^\mu x^\nu G_{\mu\nu}} \ket{0} = 
a \, e^{i \tilde{k}^\mu \tilde{x}^\nu G_{0 \mu\nu}} \ket{0}
\end{equation}
where $a \in {\cal A}_0$ is in the zero momentum subalgebra. It is then obvious that
the star product respects the tensor product structure.
\begin{equation}\label{factorization}
a_1  e^{i \tilde{k}_1^\mu \tilde{x}^\nu G_{0 \mu\nu}} \ket{0}
 * a_2  e^{i \tilde{k}_2^\mu \tilde{x}^\nu G_{0 \mu\nu}} \ket{0}=
(a_1*a_2) e^{-\frac i2 \tilde{k}_1^\mu \tilde{k}_2^\nu \theta_{0 \mu\nu}} 
e^{i (\tilde{k}_1+\tilde{k}_2)^\mu \tilde{x}^\nu G_{0 \mu\nu}} \ket{0}
\end{equation}
Recalling the structure of the BRST operator it is also obvious that after this 
rescaling in the limit $t \to \infty$ all the terms with momentum operators 
vanish and therefore it acts only on the ${\cal A}_0$ component.
In conclusion the full \sf algebra looks as 
\begin{equation}
{\cal A}={\cal A}_0 \otimes {\cal A}_1
\end{equation}
where ${\cal A}_0$ is the complicated stringy subalgebra of the string states of zero momentum in
the noncommutative directions and nonzero momentum in the commutative ones. 
The second factor $\aa_1$ is the algebra generated by the functions $e^{ikx}$ using the Moyal product.
Its precise content, K-theory and physical applications in the important cases of (compactified) 
Moyal plane and \nc torus will be our primary concern in the next section.

\section{Solutions of the \sf equations and  K-theory}

As the first case let us consider the flat Minkowski
space with $g_{\mu\nu}=\eta_{\mu\nu}$ and for simplicity assume that the rank of the B-field is two.  
This was studied in \cite{GMS, DMR, HKLM}. 
For the algebra $\aa_1$ of functions on the \nc plane let us take the Schwarz space 
${\cal S}(\rr^2)$. The associated algebra of Weyl ordered operators generates the algebra 
of the trace-class operators whose norm closure is the algebra ${\cal K}({\cal H})$  
of compact operators on a separable Hilbert space ${\cal H}$ \cite{Robert}. 
This algebra does not contain the identity, we may wish to 
add it by hand. This formally corresponds to the one point compactification of the Moyal plane. 
Thus we have up to an isomorphism
\begin{equation}\label{K+C} 
{\cal A}_1 = {\cal K} \oplus \cc {\cal I}
\end{equation}
The $K_0$ group of this algebra which will play some role later is
\begin{equation}
K_0({\cal A}_1) = \zz \oplus \zz
\end{equation}
For a general algebra it is defined as the additive group of formal differences of 
certain equivalence classes of projectors. For a detailed exposition see \cite{Wegge-Olsen}.

Let us discuss now some solutions to the string field equation of motion
in the background of large B-field. From the action (\ref{action}) it takes the form
\be\label{eom} 
Q \Psi + \Psi * \Psi =0
\ee
The basic solution is $\Psi=A_0 \otimes {\cal I}$. It is the famous solution describing the decay 
of the D25-brane which was first investigated numerically in \cite{SZ}.
The value of the \sf action per unit 
time\footnote{Throughout the whole paper we are interested in time independent configurations 
and hence the word action will always mean the action per unit time.}  
is (using the Sen's conjecture for $B=0$)
\be
S[A_0 \otimes \ii] =  2\pi \alpha' B M
\ee
in accord with the Sen's conjecture for $B$ large. Here $M$ stands for the D25-brane mass 
in the absence of any B-field 
\be
M= \frac{1}{2\pi^2} \frac{1}{\alpha' g_o^2} \int\! \sqrt{g} \, d^{25}x
\ee
The factor $2\pi \alpha' B$ comes from the effective open string coupling constant and 
from the normalization of the inner product
\be 
\bra{0,0}c_{-1}c_0 c_1 \ket{0,0} = \int\! \sqrt{G} \, d^{26}x
\ee
and accounts
precisely for the change in the mass of the D25-brane due to the background B-field. Note that there are 
some subtleties since the mass $M$ diverges. To make it finite, we should introduce some cutoff, 
which however spoils the structure of the algebra. Nevertheless the simplicity of the GMS construction
partially justifies this slightly heuristic treatment. The more careful treatment of the \nc torus will 
be given later.

As was noticed by \cite{HKLM} on the level of the low energy action and by Witten \cite{Witten:NCtach} 
from the \sft point of view one can get whole family of new solutions  
of the form $A_0 \otimes \rho$ where $\rho \in {\cal A}_1$ is any projector.  
Suppose now for a while that $\rho$ is a projector 
onto a finite $n$ dimensional subspace of ${\cal H}$. For all of these solutions one can 
easily calculate the 
value of the \sf action using the Sen's conjecture for the D25 brane without any B-field. 
Let us list some of them in the suggestive form
\begin{displaymath}
\begin{array}{|l|r|c|}
\hline
{\rm Solution} & {\rm Value \ of \ the \ action} & {\rm Interpretation} \\ \hline 
A_0 \otimes \ii &  2\pi \alpha' B M & D25 \to vac \\ \hline
A_0 \otimes (\ii-\rho) &  \left(2\pi \alpha' B -n \frac{\alpha'}{R^2} \right) M & D25 \to n D23 \\ \hline
A_0 \otimes \rho & n \frac{\alpha'}{R^2} M & n D23 \to vac \\ \hline
\end{array}
\end{displaymath}
In order to get finite results we had to regularize the area of the Moyal plane 
(in closed string metric) to be $(2\pi R)^2$. We also need the formula 
$\int d^2x \, \rho(x)=2\pi\theta n$ from \cite{GMS}. 
The values of the action for the above solutions exactly correspond to the decay
energies between various D-brane systems. This leads to the interpretation listed in the last column.
One can readily enlarge the above table to include various lower dimensional branes and also to 
introduce Chan-Paton factors introducing thus more D25-branes.

Sum of any two solutions (or two projectors) is a solution (or a projector) only if the two projectors 
are orthogonal. To be able to add any two projectors K-theory does what in physical terms is called 
introducing Chan-Paton factors. At this point it should be clear that we can also interpret 
the $K_0$ elements as formal differences of branes. The \sf action clearly acts as a homomorphism on 
this group. This interpretation of $K_0$ is somehow reminiscent of the situation in IIB theory 
\cite{Witten:Ktheory} where the elements are formal differences of {\it bundles} 
on branes and antibranes. Nonetheless, the fact that it is a completely different group
is more than obvious. It may seem that the K-theory we are talking about is just an artefact
of the 'hand-made' algebra (\ref{K+C}). By looking at the example of the (more realistic) \nc torus 
we shall try to convince the reader that there is something deeper going on.

\subsection*{Noncommutative torus} 

Let us briefly discuss the case of the \nc torus. Here the relevant algebra $\aa_1$ is the well known 
rotational algebra $A_\theta$. Its $K_0$ group is the same as for the compactified Moyal plane above. 
Unfortunately in this case the beautiful construction \cite{GMS} of all the 
projectors breaks down even though one still has a homomorphism from the algebra $\aa_1$ to the space
of bounded operators by an analog of the Weyl quantization formula. 
Some representatives of all the equivalence classes 
of projectors were nevertheless constructed by Rieffel \cite{Rieffel}. 
The Powers-Rieffel projector on the torus $[0,2\pi]^2$ takes the form 
(in the representation by ordinary functions)
\be
p(x_1,x_2)=2\cos(x_1) g\left(e^{i(x_2+\frac{\theta}{2})}\right) + f\left(e^{ix_2}\right)
\ee
where $f$ and $g$ are two functions satisfying certain relations. These can be chosen to be sufficiently 
smooth if one wishes. The trace on the \nc algebra $\aa_1$ in the representation by ordinary functions
with the Moyal product is just an ordinary integral over the torus (normalized by the total area)
which gives precisely $\frac{\theta}{2\pi}$.
From our point of view the only problematic feature of these solutions is that they are not 
well localized in one direction (in this case $x_1$). This prevents us from looking at those solutions 
as codimension two lump representing lower dimensional brane. Nevertheless from our experience with 
the Moyal plane, we believe that there should exist also well localized solutions with straightforward 
physical interpretation. 

General theorem due to Pimsner and Voiculescu when combined with the Rieffel's construction 
\cite{ Wegge-Olsen,Rieffel} 
states that the range of the trace on 
projections in $\aa_1$ is exactly $(\zz+\frac{\theta}{2\pi} \zz) \cap [0,1]$. 
The unusual normalization factor $\frac{1}{2\pi}$ comes from requiring the standard form
(\ref{factorization}) of the star product.
Calculating the \sf action
for the solution $A_0 \otimes p$ one gets
\be
S[A_0 \otimes p] = 2\pi \alpha' B M \Tr p = 2\pi \alpha' B M \left(m-\frac{\theta}{2\pi}n\right)
\ee
where $m,n \in \zz$ are such that
\be
m-\frac{\theta}{2\pi}n \in [0,1]
\ee
We see that for $m=1$ and $n \in \nn$ not too large (such that the projector exists)
we get {\it precisely} the same values as those for the Moyal plane above.
It is perhaps curious to note that the theorem also asserts that even without 
introducing the Chan-Paton factors one can describe the decay of 
$m>1$ D25 branes into an appropriate number of D23 branes. 
This is not true for the Moyal plane case.

As we said above one may have doubts about the role of K-theory on the Moyal plane.  
But here on the \nc torus in order to find a single example of a projector we 
had to use the K-theoretical sources. Strikingly these projectors lead to the 
correct masses of D-branes, exactly as the GMS projectors. 
The projectors in \ncg are primarily used to define
projective modules --- a \nc generalization of vector bundles --- which are naturally 
classified by K-theory.
To end up this section we would like to make the following interesting remark.
The \sf action is the (secondary) Chern-Simons class of the \nc bundle defined by the connection which
is the string field. In the large B-field limit
when the algebra factorizes the action becomes equal up to a factor to the Chern class of 
a completely different \nc bundle over the torus specified by the choice of the 
projector $p$. We believe that further investigations may reveal beautiful interplay 
between these objects in \nc geometry.

\section{Proposal for the exact solution of the tachyon potential}

The proposal is based on the following simple observation: All the decays of D25 brane that 
are described in the large B-field limit by taking a nonzero projector $\rho \in \aa_1$ 
are related to each other by a nonunitary isometry\footnote{An operator $U$ for which $U^\da U$ 
is projector is called a partial isometry. Then automatically $U U^\da$ is a projector. 
If $U^\da U=\ii$ then $U$ is called an isometry.}.
Of course it doesn't mean that they are in the same 
K-theory class since this isometry doesn't belong to $\aa_1$. To give an example
consider the solutions describing the decays $D25 \to vac$ and $D25 \to (n) D23$ with projectors 
$\ii$ and $\ii-\rho$ respectively.  

The isometry $U$ which relates them as follows
\bea\label{pi}
\ii-\rho &=& U  U^\da   \nonumber\\
\ii &=& U^\da U
\eea
can be found in some cases explicitly. 
If for instance we take $\rho=\ket{0}+\ket{1}+\cdots+\ket{n-1}$ then $U$ and $U^\da$ are the 
ordinary shift operators
\bea
U &=& \sum_{m=0}^{\infty} \ket{m+n} \bra{m} \nonumber\\
U^\da &=& \sum_{m=0}^{\infty} \ket{m} \bra{m+n} \nonumber\\
\eea
The operator $U$ is clearly noncompact (and it is neither unity) so it does not belong to $\aa_1$. 
Thus the projectors $\ii$ and $\ii-\rho$ do not have to belong to the same K-theory class. They 
would, however, if we were working with the algebra of all bounded operators. 

Note that \sf solutions representing the above decays are related by a formula 
which formally looks like a \sf gauge transformation
\be
A_0 \otimes (\ii-\rho) = U (Q+ A_0 \otimes \ii ) U^\da
\ee
The first term on the right hand side gives of course zero contribution since $Q$ doesn't 
act on $\aa_1$. More useful relation is obtained by multiplying with $U^\da$ and $U$ on the left 
and right respectively
\be
A_0 \otimes \ii = U^\da (Q+ A_0 \otimes (\ii-\rho) ) U
\ee
Our conjecture is as follows: Since the decays $D25 \to vac$, $D25 \to n D23$ and so on are related 
by gauge-like isometry transformation, it is natural to expect that in the full string theory also the 
trivial process $D25 \to D25$ described by the zero string field is related to the others 
in a similar way. Thus we expect
\be\label{guess}
A_0 = V^\da * Q V
\ee
for some $V$ acting on $\aa_0$ and satisfying 
\bea\label{conditions}
V^\da * V &=&\ii \nonumber\\
Q(V *V^\da) &=&0
\eea
where the star is now the stringy product (not the Moyal one) and the dagger means the usual 
star involution of the \sf algebra.
The last equation (which is of course also satisfied by $U$) was added in order 
to fulfill the equation of motion. Note that both conditions (\ref{conditions}) could 
be replaced simply by $V*V^\da=\ii$ but this is not favored by our analogy.  
It is straightforward to check the \sf equations of motion (\ref{eom}) provided one can use 
the associativity of the algebra. This is however not a priori clear since $V$ 
(in analogy with $U$) doesn't appear to be an element of the algebra. 
Indeed it is well known that when one tries to add some elements to the so far not properly 
defined string algebra one runs into problems with associativity anomalies \cite{Horowitz,Rastelli}. 
We end up by emphasizing the importance of clarification of these anomalies and 
by expressing the hope that it will be soon possible to confirm the above conjecture at least
numerically.

\section*{Acknowledgements}
I would like to thank Thomas Krajewski and Alessandro Tomasiello for numerous valuable
discussions and to Loriano Bonora for critical reading of the manuscript. I thank also
Martin O'Loughlin for useful comments and to Gianluca Panati for looking up the reference
\cite{Robert}.
Thanks also to Barton Zwiebach whose lectures on the ICTP Spring Workshop on Superstrings
initiated my interest in the subject. 

\end{document}